%% file: main.tex
\documentclass{iopart}
\usepackage[label font=bf,labelformat=simple]{subfig}
\usepackage{floatrow}
\usepackage{graphicx}
\usepackage{caption}
\usepackage{amssymb}
\begin{document}

\title{Vibration Isolation for the Laser Interferometer Lunar Antenna}

\author{Brett N. Shapiro$^1$}

\address{$^1$Space Exploration Sector, The Johns Hopkins University Applied Physics Laboratory, Laurel, MD}

\ead{brett.shapiro@jhuapl.edu}

\begin{abstract}
The Laser Interferometer Lunar Antenna (LILA) presents a novel concept for observing gravitational waves from astrophysical sources at sub-Hertz frequencies. Compared to the Earth, the seismic environment of the moon, while uncertain, is known to be orders of magnitude lower, opening the possibility for achieving this sub-Hz band. This band fills the gap between space-based detectors (mHz) and Earth-based detectors (10 Hz to a few kHz). The initial version of LILA, known as LILA Pioneer, calls for non-suspended optics, relying on the moon's resonant modes to respond to gravitational waves. However, the follow-on design, LILA Horizon, requires suspensions to realize in-band free floating test masses and to filter the residual seismic background. This paper will establish baseline designs for these suspensions for different assumptions of the seismic background.
\end{abstract}

\input{intro.tex}
\input{susdesign.tex}
\input{risks.tex}
\input{conclusion.tex}

\section*{Acknowledgments}
Thanks to Karan Jani for the input on the science case, to Valerio Boschi and Massimiliano Razzano for the discussion on IPs and anti-springs, and to Matt Abernathy for discussions about coating brownian noise and suspension dilution factors. This work was supported by the Space Formulation Mission Area at the Johns Hopkins Applied Physics Laboratory.

\section*{References}
\bibliography{Refs} 

\newpage
\input{appendix.tex}

\end{document}

%% file: intro.tex
\section{Introduction}

The moon presents a unique opportunity and environment for gravitational wave observation. The low seismic noise floor makes it possible to create a detection band from approximately 0.1 Hz to 10 Hz, which fills the gap between Earth-based and space-based detectors. The Laser Interferometer Lunar Antenna (LILA) aims to construct a moon-based observatory to exploit this gap. While the initial incarnation of LILA, LILA Pioneer, will have optics fixed to the ground for observing the moon's resonant modes, the follow-on observatory, LILA Horizon will utilize test mass suspensions \cite{LILApaper, LILA_WP}.

\begin{figure}
 \centering
        \includegraphics[width=4in]{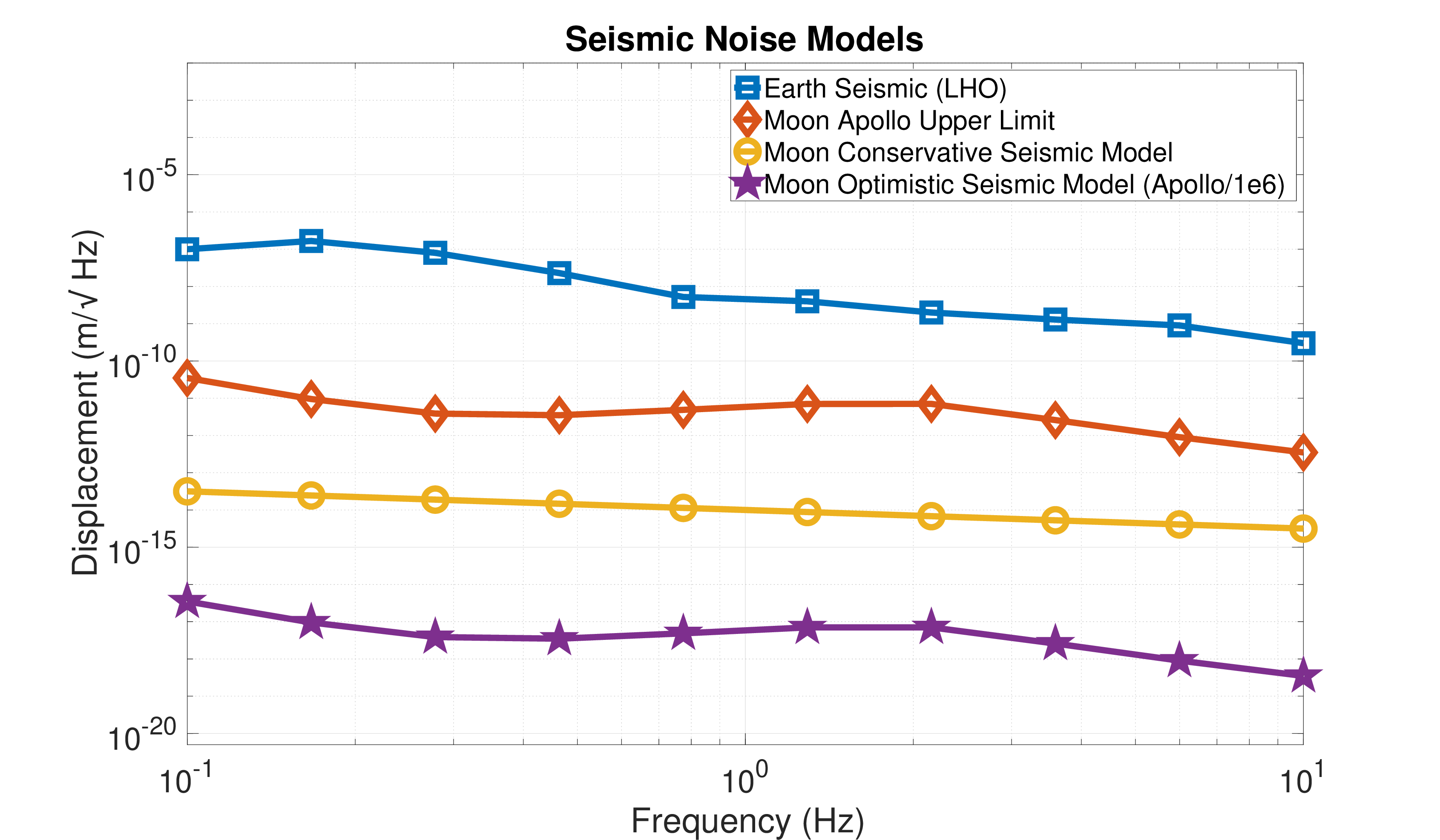}
 \caption{A summary of seismic noise curves. The blue square curve is an example of seismic noise on the Earth as observed at the LIGO Hanford Observatory \cite{LHOseis}. The red diamond curve is the moon's upper limit given by the noise floor of the sensors placed by the Apollo missions \cite{moonseis}. The yellow circle curve is the estimated moon environment given by \cite{Cozzumbo}. The purple star curve is an optimistic estimate given by reducing the Apollo sensor noise floor by 1 million.}
\label{fig:seismic}
\end{figure}

Figure \ref{fig:seismic} shows estimates of the moon's background seismic noise \cite{Cozzumbo, moonseis} in comparison to a representative Earth measurement \cite{LHOseis}. The moon's background is believed to be considerably less than the red diamond upper limit imposed by the Apollo seismometers \cite{moonseis}. The uncertainty in how much less spans some orders of magnitude. The yellow circle curve is a reasonable representative guess \cite{Cozzumbo}. The purple star curve, labeled as the optimistic model, is set by dividing the Apollo upper limit by a million. This scaling is rather arbitrary. However it is a useful curve in the sense that it permits exploration of a regime where suspension designs are thermal noise limited across the entire detection band rather than seismic noise limited in part.  

This paper presents in Section \ref{sec:susdesign} an initial exploration of the mechanical parameter space for room temperature, 290 K, suspension conceptual designs for LILA Horizon. These designs focus on simplicity and robustness while still justifying the science case. Future work may consider more advanced and higher performing designs, which can be compared against the results here. Section \ref{sec:risks} lists the known risks in realizing such room temperature suspensions on the moon. Cryogenic lunar suspensions could be a useful approach for reducing thermal noise; however, room temperature technology is much more mature, so this paper is focused on that regime. 

%% file: susdesign.tex
\section{Suspension Design and Performance}
\label{sec:susdesign}

The overall goal in suspension design is to adequately filter seismic disturbances while not contributing to other noise sources. This goal decomposes into various requirements.
\begin{enumerate}
\item The test mass seismic motion is sufficiently below the observatory design sensitivity.
\label{req1}
\item The suspension thermal noise is sufficiently below the observatory design sensitivity.
\label{req2}
\item The rigid body resonant frequencies are sufficiently outside the detection band.
\label{req3}
\item The flexible body resonant frequencies such as the wire violin modes are sufficiently above the detection band.
\label{req4}
\end{enumerate}
The above requirements apply directly to both the horizontal axis aligned with the detector arms and the vertical degrees of freedom, but also indirectly to all other degrees of freedom given that coupling is likely to exist. The coupling from vertical to horizontal is fundamental at approximately $2.3\%$ given the non-negligible 40 km detector arm length with respect to the moon's 1737 km radius. That is, a local vertical oscillation of an optic at one end of an arm has a horizontal component with respect to the optic at the other end. In this paper we will only consider performance along the horizontal and vertical degrees of freedom. The performance of the other degrees of freedom will be left for future work.

In all the figures showing strain noise below, the suspension noise is scaled up by $\sqrt{2}$ to account for two end test mass suspensions in the LILA Horizon design. The number of suspensions may need to be updated in future analyses as the LILA design evolves. 

The material properties used in the calculations of suspension performance are given in \ref{subsec:matprop}. Thermal noise calculations rely on applications of the Fluctuation Dissipation Theorem shown in \ref{subsec:FDT}, which are modeled with GWINC \cite{pygwinc}, adapted for the moon environment.

\subsection{Optimistic Seismic Noise Assumption}
\begin{figure}
 \centering
        \includegraphics[width=1.5in]{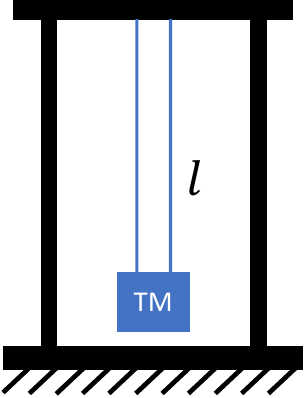}
 \caption{A simple one-stage room temperature suspension model of length $l$. The 100 kg test mass is supported by four silica glass fibers. There are no cantilever springs for vertical isolation.}
\label{fig:sus_1stage}
\end{figure}

The first candidate designs begin with the simplest assumption, that the seismic noise follows the optimistic curve shown in Figure \ref{fig:seismic}. This assumption permits the consideration of a single stage room temperature suspension with no cantilever springs, as shown by Figure \ref{fig:sus_1stage}.

This single stage suspension has a test mass of 100 kg and, like Advanced LIGO, is supported by four silica glass fibers \cite{Cumming_2012}. Silica fibers are used due to their advantageous thermal noise properties at room temperature. The very small internal mechanical damping focuses much of the thermal noise at the resonant frequencies, leaving it lower off-resonance. As will be seen, thermal noise is one of the main limitations for these suspension designs.

\begin{figure}
 \centering
        \includegraphics[width=5.5in]{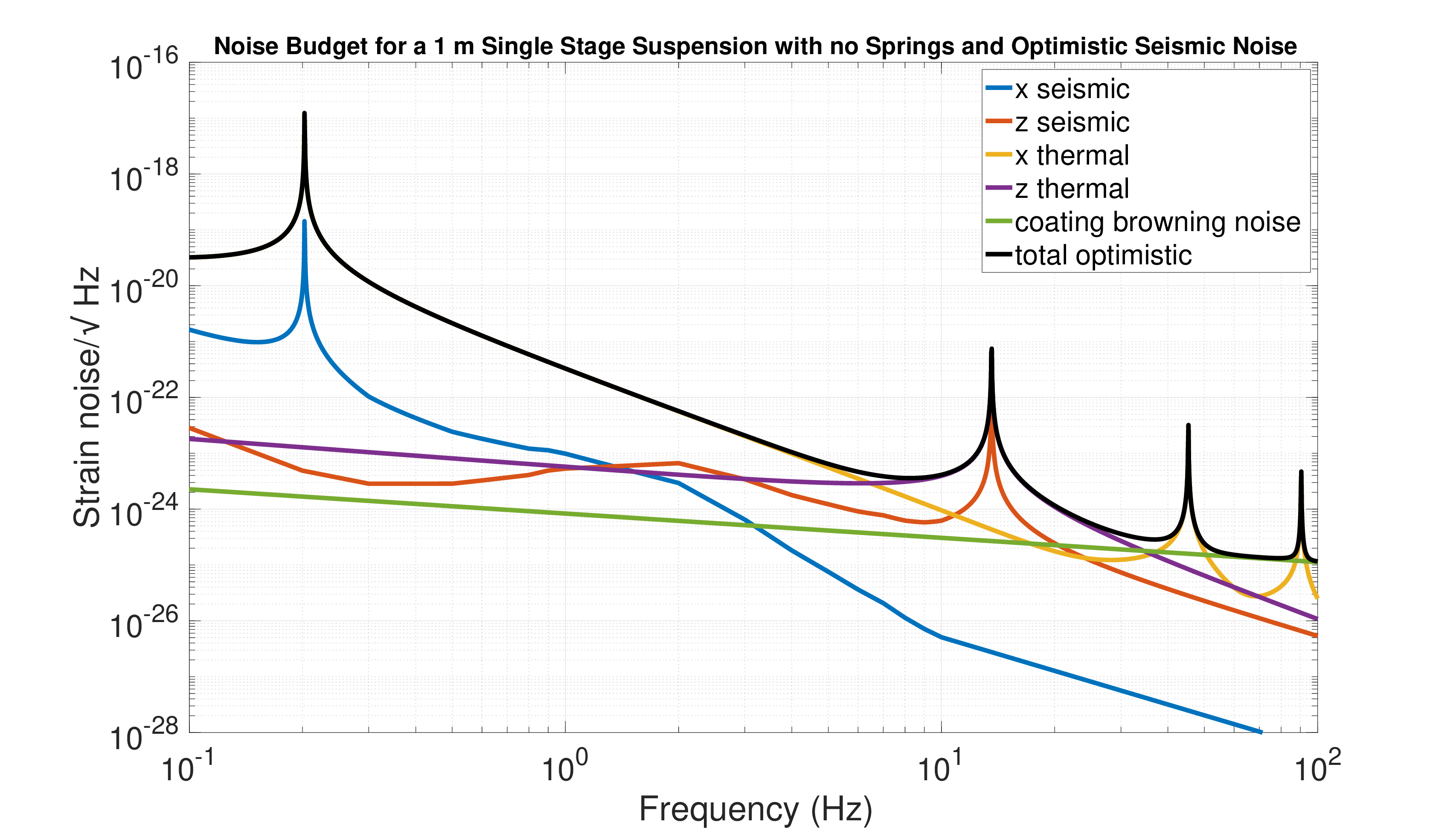}
 \caption{Seismic and thermal strain noise budget with the optimistic seismic assumption. The suspensions are room temperature, one-stage, 1 m long, 100 kg suspended, no springs, and silica fibers stressed to 186 MPa in the bending region to cancel thermoelastic stress and 16 MPa along the majority of the length to push the vertical bounce mode up and out of band. The suspension architecture is given by Figure \ref{fig:sus_1stage}. Performance is limited entirely by thermal noise in the 0.1 Hz to 10 Hz band, horizontal at low frequencies and vertical at higher frequencies. Coating brownian noise is given by coatings for LIGO A+ \cite{AplusNoise} and a 50 cm diameter optic.}
\label{fig:NoiseBudget_1m_SingleStage_NoSprings_LowStress}
\end{figure}

Figure \ref{fig:NoiseBudget_1m_SingleStage_NoSprings_LowStress} shows the gravitational wave strain noise budget for a 1 m long single suspension in LILA Horizon. The fiber stress is set to 186 MPa in the bending regions at the top and bottom to cancel the thermoelastic component of thermal noise. Along the majority of the length of the fiber, the stress is set to just 16 MPa to push the vertical bounce mode above 10 Hz so that it is out of band. For reference, the Advanced LIGO fiber stress is 800 MPa \cite{Cumming_2012} and the A+ stress is planned to be 1200 MPa \cite{AplusFibers}. The max stress here, in the skinny bending region, is more than 4 times lower than the max Advanced LIGO stress. This lower max stress will help with the mechanical robustness of the fibers, given that regular repairs will not be an option should one break. Robustness also helps survival within the vibratory environments of launch and landing. Testing will be required to determine if these fibers are sufficiently robust. This stress profile results in a fiber radius profile of 263 $\mu m$ in the bending regions and 898 $\mu m$ in the longer center region.

\begin{figure}
 \centering
        \includegraphics[width=5.5in]{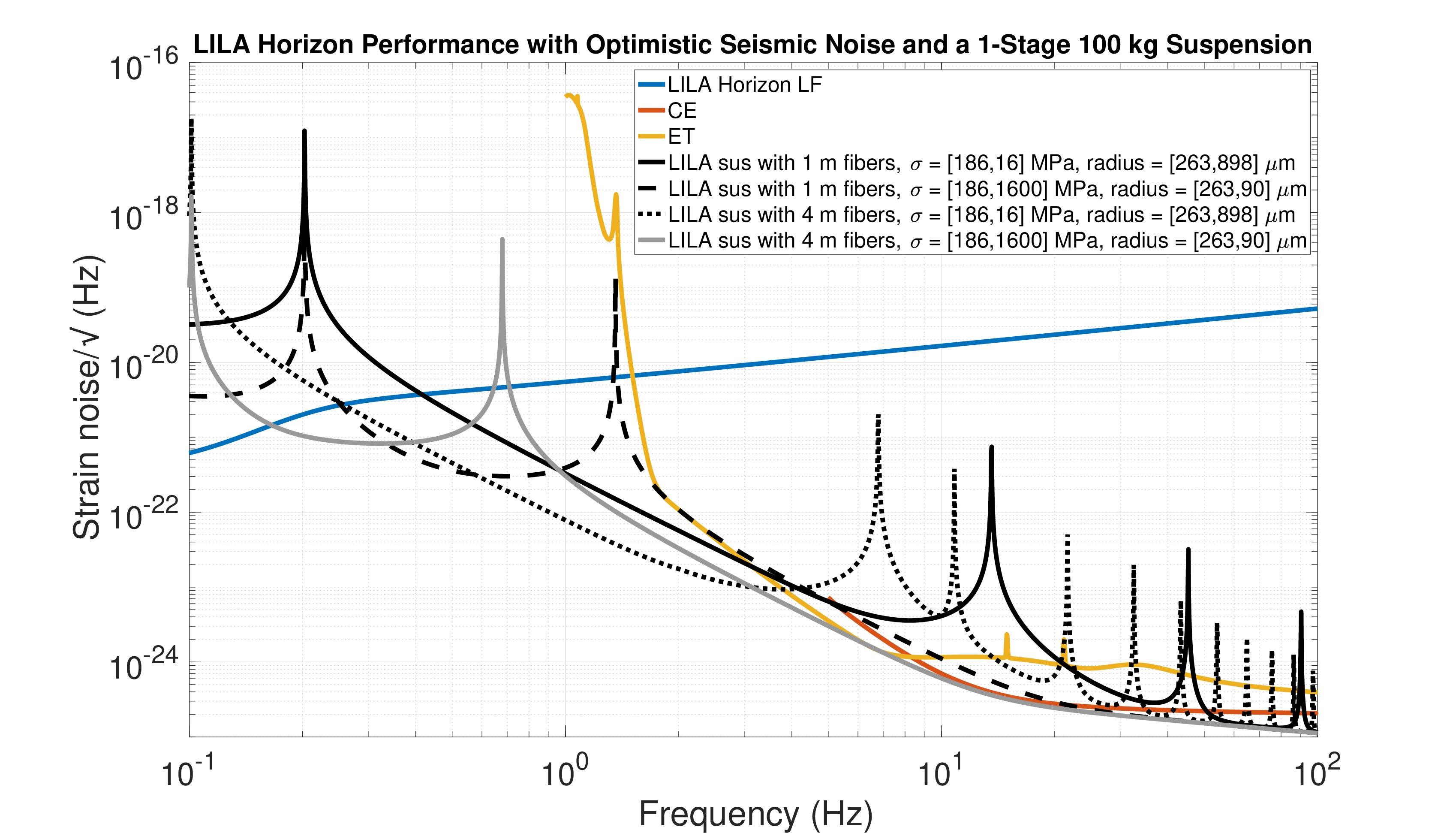}
 \caption{Total LILA suspension strain noise, for 100 kg one-stage suspensions with optimistic seismic, compared to LILA Horizon LF \cite{LILApaper}, ET \cite{ETnoise}, and CE \cite{CEnoise}.  A few variations in suspension design are presented. The peak above 10 Hz in the 4 m dotted black curve is due to the fundamental violin mode of the silica fibers.}
\label{fig:single_otpimistic_sus_total}
\end{figure}

Figure \ref{fig:single_otpimistic_sus_total} shows the total strain noise performance of the suspension given by Figure \ref{fig:NoiseBudget_1m_SingleStage_NoSprings_LowStress} in solid black against the low-frequency component of the LILA Horizon observatory, LILA Horizon Low Frequency (LF) \cite{LILApaper}; and the proposed third generation Earth-based observatories Cosmic Explorer (CE) \cite{CEnoise} and the Einstein Telescope (ET) \cite{ETnoise}. The figure also shows three other variations of the suspension design, with either 1 m or 4 m long fibers and either high or low stress fibers. The two fiber stress and radius numbers shown in the legend for the four LILA cases represent the values for the skinny bending regions and the wider central regions respectively. The high stress case is with fibers stressed to 1600 MPa, twice the Advanced LIGO fiber stress. These high stress cases are meant to be illustrative of whether pushing design limits can help in any way rather than to be suggestive of alternate practical solutions. Indeed, the minimum breaking stress, according to \cite{AplusFibers}, is 2000 MPa, only 25\% higher. 

The four design variations for this optimistic seismic assumption illustrate that frequency dependent trade-offs in performance exist as a function of length and stress. First, longer lengths help at lower frequencies. Increasing the length however lowers the bounce and violin mode frequencies, which contribute noise to the higher end of the detection band. Second, fiber stress also affects what frequencies are most sensitive. Higher stress improves lower frequencies because thermal noise from the violin modes are moved out to higher frequencies. However, the bounce mode frequency moves down, invading the higher end of the detection band. 

Setting the bounce mode lower than the detection band would be impractical given that the fibers cannot tolerate the stress necessary to do this, necessitating silica springs. Research and development would be required for such springs. 

Notably, low stress fibers invert the fiber profile relative to Advanced LIGO. Here we have thinner ends and a wider middle section. LIGO has wider ends and a thinner middle section. This inversion is because low stress pushes the bounce mode beyond the LILA 10 Hz high frequency end of the detection band while LIGO pushes it below the 10 Hz low frequency end. 

\begin{figure}
 \centering
        \includegraphics[width=5.5in]{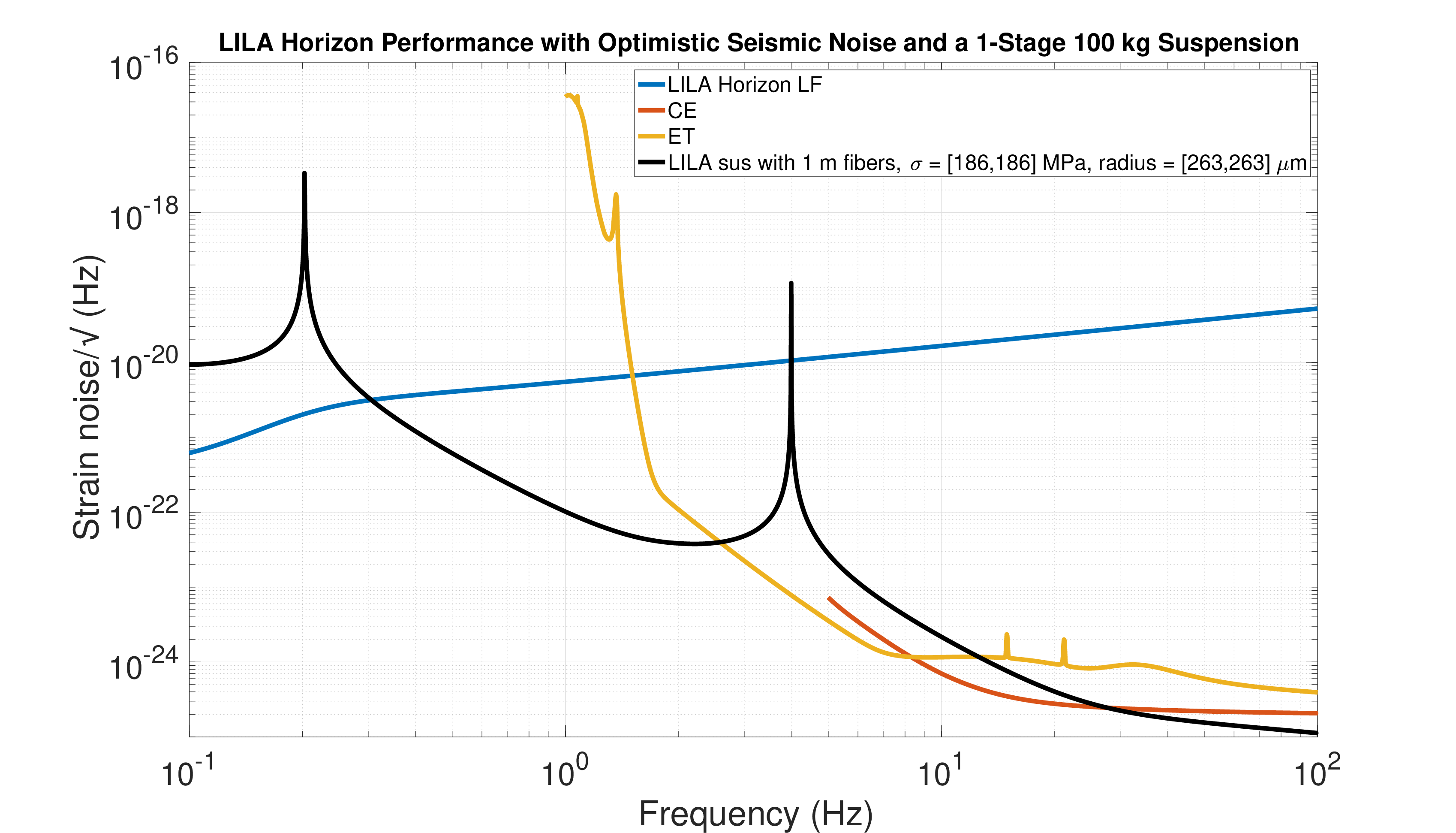}
 \caption{Total LILA suspension strain noise, for a possible baseline suspension design in the optimistic seismic case: a 100 kg, one-stage suspension with silica fiber stress set to a uniform 186 MPa along the entire length. Its performance is compared to LILA Horizon LF \cite{LILApaper}, ET \cite{ETnoise}, and CE \cite{CEnoise}. }
\label{fig:single_baseline_otpimistic_sus_total}
\end{figure}

A possible baseline design in this optimistic case is a 1 m, 100 kg, single stage suspension with a uniform 186 MPa stress set throughout the fiber length. This would result in a simpler fiber that does not need to change radius along the length. The bounce mode moves below 10 Hz into what might be still be considered part of the detection band, but it is in a region where performance is otherwise no better than proposed Earth-based observatories. The stress is also still low enough that the fibers are testable in Earth gravity, at stresses similar to the A+ design (1.2 GPa). Figure \ref{fig:single_baseline_otpimistic_sus_total} shows the performance of this design. 

It is worth noting that both the horizontal and vertical suspension thermal noises scale by $\frac{1}{\sqrt{m}}$, where $m$ is the test mass mass, according to Equation \ref{eq:single_sus_therm} in \ref{subsec:FDT} (assuming $k$ must scale with $m$). Since performance in this optimistic seismic case is limited entirely by suspension thermal noise, performance can be improved by increasing the mass, while increasing the fiber diameter to maintain the stress, until seismic noise dominates or practical limitations are reached. 

%%%%%%%%%%%
\subsection{Conservative Seismic Noise Assumption}
\begin{figure}
 \centering
        \includegraphics[width=5.5in]{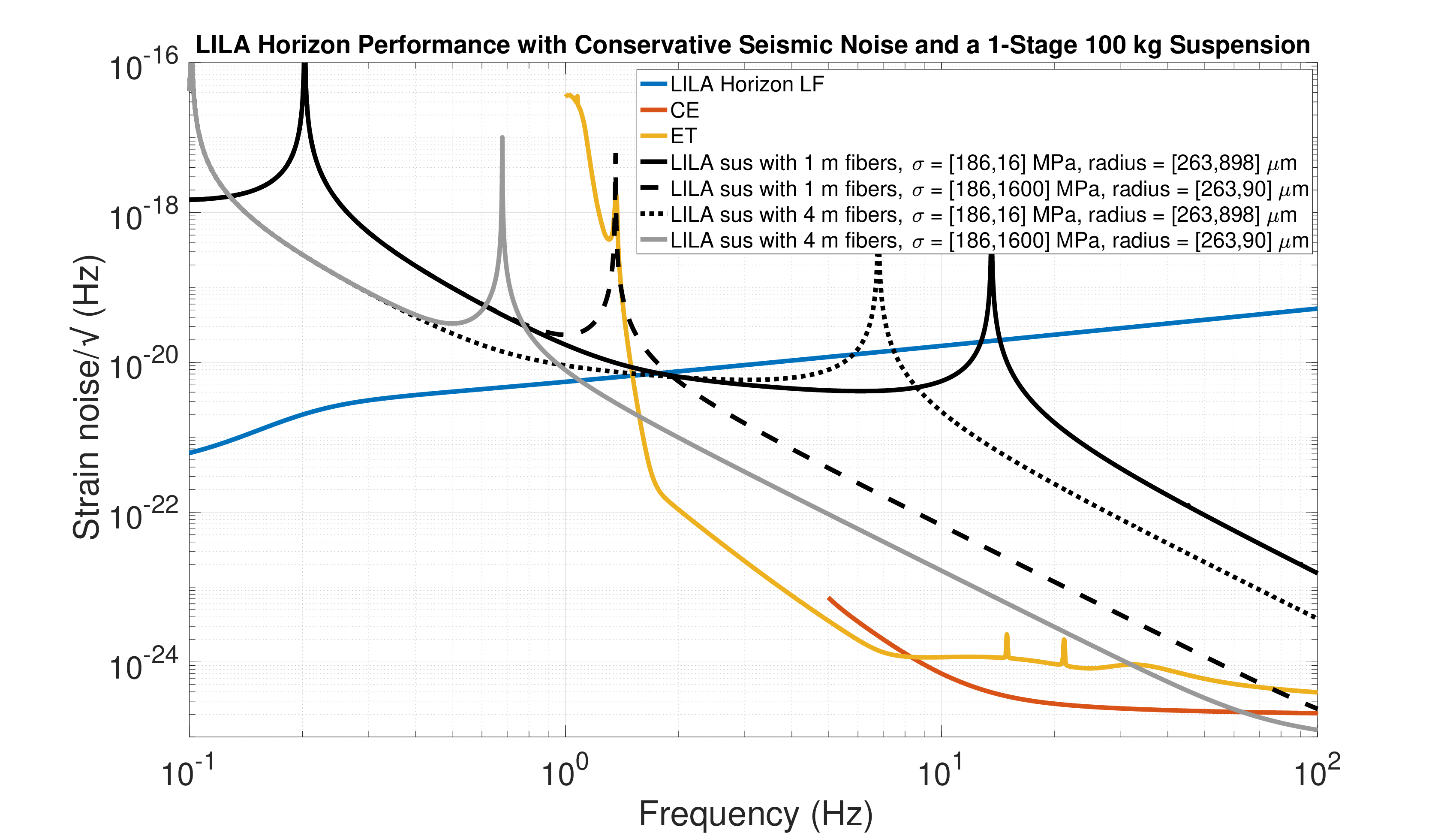}
 \caption{Total LILA suspension strain noise, for one-stage suspensions with conservative seismic, compared to LILA Horizon LF \cite{LILApaper}, ET \cite{ETnoise}, and CE \cite{CEnoise}.  A few variations in design are presented.}
\label{fig:single_conservative_sus_total}
\end{figure}

The next set of candidate suspension designs consider the conservative seismic curve shown in Figure \ref{fig:seismic}. If we use the same simple one-stage suspension designs given in Figure \ref{fig:single_otpimistic_sus_total}, then performance is much worse, as shown in Figure \ref{fig:single_conservative_sus_total}. Only the 4 m long suspension with high stress fibers beats both LILA Horizon LF and ET, and only slightly from 1.1 Hz to 1.6 Hz. Unlike the optimistic case, these suspensions are seismic noise limited rather than thermal noise limited; and unlike thermal noise, increasing the mass of a single stage suspension has no impact on seismic performance. More complex suspension designs are required if this seismic situation is considered.

Filtering this seismic noise in the horizontal direction requires some combination of longer lengths, inverted pendulums (IPs), and more stages. Filtering the seismic noise in the vertical direction requires springs. To get vertical modes low enough, those springs need to be anti-springs. Anti-springs have historically been constructed in two different ways in the gravitational wave community, either with geometric anti-springs \cite{GASprings} or with magnetic anti-springs \cite{VirgoSus, VirgoMAS}. Either variety may be considered here.

\begin{figure}[htb]\centering
\sidesubfloat[]{\includegraphics[width=1.5in]{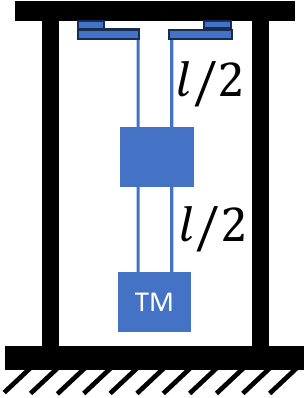}\label{fig:a}}
\hfil
\sidesubfloat[]{\includegraphics[width=1.5in]{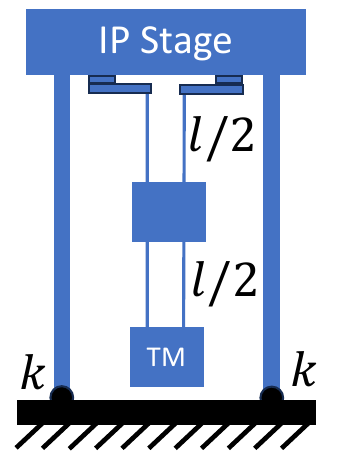}\label{fig:b}}
\caption{Two conceptual multi-stage suspension designs for the high seismic noise assumption. Subfigure (a) is a two-stage concept, with vertical springs at the top stage. Subfigure (b) is similar except that the base supporting the springs is also a massive stage, supported by IP legs. Thus concept (b) is three stages in the horizontal but only two in the vertical. In both concepts, the springs utilize anti-springs to reduce the resonant frequency.}
\label{fig:2and3stagesus}
\end{figure}
% directions for the subfigures came from here: https://tex.stackexchange.com/questions/310081/subfigures-to-add-a-and-b-to-subfigures-in-the-top-left-corner-and-to-label

When considering multiple stages, the work of \cite{Edgard} leads to the following design rules of thumb.
\begin{itemize}
\item Number of stages: Increasing the number of stages increases the highest rigid body resonant frequency. At some point, adding stages won’t help isolation because the modes will spread out and invade the detection band.
\item Mass distribution: The optimal mass distribution locates more mass at higher stages. This optimization is non-trivial only for three or more stages, given a fixed total mass and a fixed test mass size. 
\item Length distribution: Longitudinal isolation is maximized by equal wire lengths at each stage.
\end{itemize}

Figure \ref{fig:2and3stagesus} shows two conceptual schematics of multi-stage suspension designs. Subfigure (a) shows a two stage design with the upper stage supported by cantilever maraging steel springs. The test mass is supported by silica fibers like in the previous section. The penultimate mass is supported by metal wires. Subfigure (b) shows a similar design except that the support stage is also a mass supported by IP legs. The IP springs are constructed of maraging steel. Thus, subfigure (b) shows three stages in the horizontal direction and two in the vertical direction. 

In both cases of Figure \ref{fig:2and3stagesus}, the mechanical loss in the metal springs is much higher than the silica fibers supporting the test mass, so thermal noise will again be a concern. The analysis below will show that swapping the metal wires and springs for silica may be desirable for room temperature suspensions. To the author's knowledge, no silica anti-springs or IP flexures yet exist. If not, technology development will be required. 

\begin{figure}
 \centering
        \includegraphics[width=5.5in]{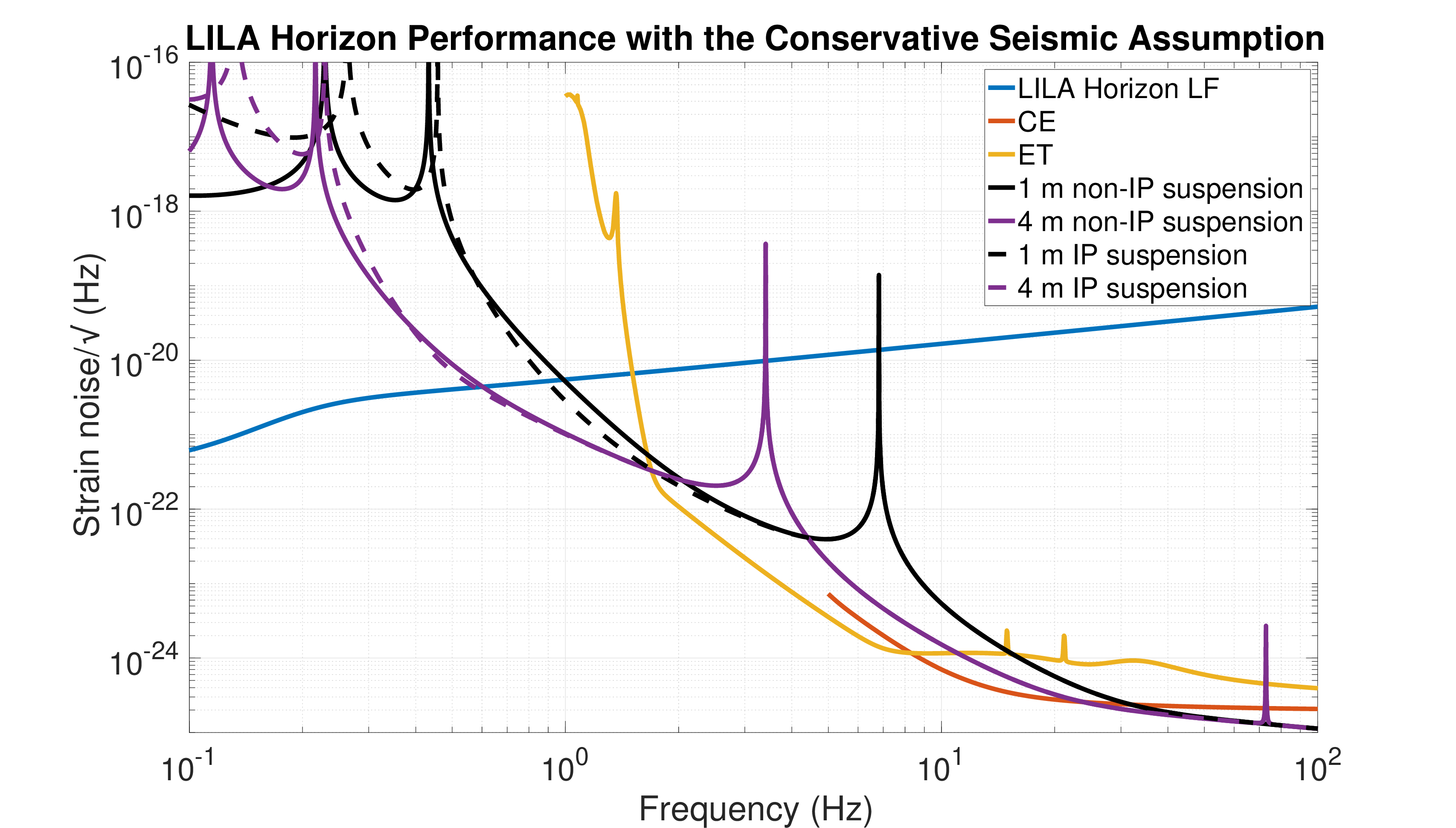}
 \caption{Total LILA suspension strain noise with conservative seismic, compared to LILA Horizon LF \cite{LILApaper}, ET \cite{ETnoise}, and CE \cite{CEnoise}. Solid suspension curves represent two-stages with 0.05 Hz metal anti-springs but no IP; dashed lines hang the same two-stage system from a 684 kg IP tuned to 0.04 Hz. Black lines are 1 m in total length, purple are 4 m. The test mass is 100 kg and the penultimate mass is 216 kg.}
\label{fig:ConservativeSusMetal_Noise_vs_3G}
\end{figure}

Figure \ref{fig:ConservativeSusMetal_Noise_vs_3G} shows the total noise for two lengths of a two-stage pendulum with and without a 40 mHz IP stage (dashed lines and solid lines respectively). In all cases, the vertical anti-springs are tuned to 50 mHz. The IP here has little benefit because the IP springs and vertical anti-springs have relatively large loss factors of 0.001 due to construction out of maraging steel. Black curves are 1 m total suspension length, purple lines are 4 m. The test mass is 100 kg and the penultimate mass is 216 kg, the IP stage is 684 kg. The silica fibers stress is set to 186 MPa to permit constant cross-section along the length. The metal wire stress is set to 103 MPa, 6 times lower than on Earth so that they are similarly strong enough for Earth gravity like the fibers.

\begin{figure}
 \centering
        \includegraphics[width=5.5in]{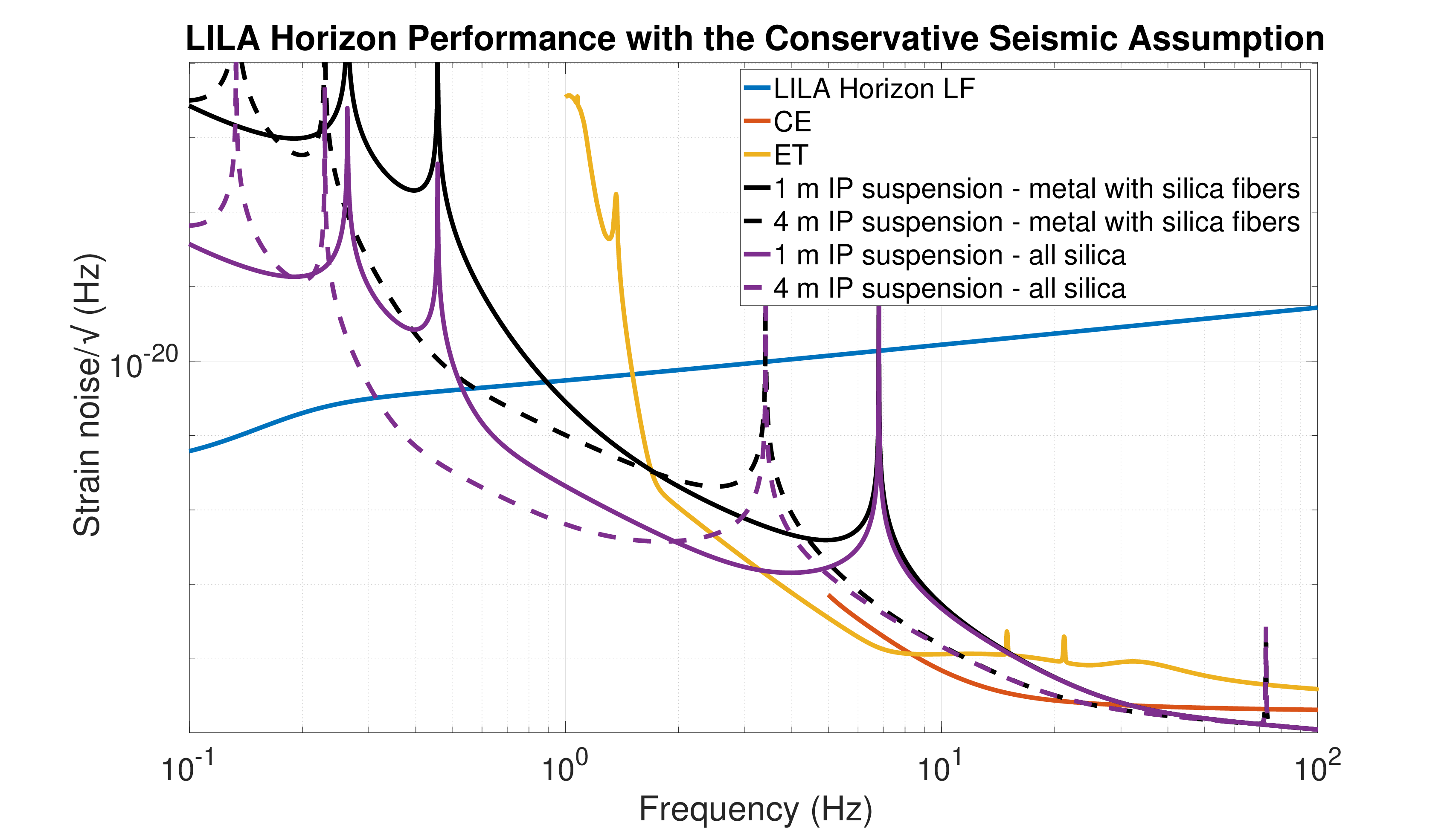}
 \caption{Total LILA suspension strain noise for two-stage suspensions supported by a 40 mHz IP with conservative seismic, compared to LILA Horizon LF \cite{LILApaper}, ET \cite{ETnoise}, and CE \cite{CEnoise}.  Black lines represent metal suspensions with silica test mass fibers, the same as the IP curves in Figure \ref{fig:ConservativeSusMetal_Noise_vs_3G}. Purple lines are the same suspensions but with all silica, including the penultimate mass wires, the anti-springs, and the IP springs. The springs have loss factors of 1e-7. Solid lines are 1 m total length, dashed are 4 m.}
\label{fig:IPsus_allglass_vs_silicafiberonly}
\end{figure}

Figure \ref{fig:IPsus_allglass_vs_silicafiberonly} illustrates how much improvement there might be if the metal springs and wires are exchanged for silica. It shows the same IP curves as Figure \ref{fig:ConservativeSusMetal_Noise_vs_3G}, but compares them to a cases where all suspension flexures are silica. While the silica fibers have loss factors on the order of 1e-8, the silica anti-springs and IP springs are set to 1e-7 for some conservativeness. 

Note, the IP and anti-spring loss $\phi$ will be magnified by the anti-restoring forces, analogous to the reciprocal of a fiber's dilution factor (see \cite{Cumming_2012} for a discussion of dilution factors). As shown by Eq \ref{eq:reverse_dilution}, this magnification factor is equal to the ratio of the spring stiffness without the anti-spring, $k_0$, to the net stiffness with the anti-spring, $k_{net}$. That is, if the anti-spring reduces the effective stiffness by 2, the loss magnification is equal to 2. Similarly, the magnification factor is equal to the square root of the ratio of suspension resonant frequency without the anti-restoring force, $f_0$, to the net resonant frequency with the anti-restoring force, $f_{net}$.

\begin{equation}
\phi_{net} = \phi_0\frac{k_0}{k_{net}} = \phi_0\sqrt{\frac{f_0}{f_{net}}}
\label{eq:reverse_dilution}
\end{equation}

More mass may improve performance here as well. The suspension with the IP, in these simulations, is at least partially thermal noise limited, so more massive test masses help suppress those noise components. More total suspended payload helps seismic noise as well. Even with a fixed test mass size, horizontal seismic isolation benefits from heavier upper stages.

%% file: risks.tex
\section{Risks to Mitigate}
\label{sec:risks}

A number of risks must be mitigated to realize successful room temperature suspension systems on the moon. This section describes the known risks as of writing this article, but other risks may exist.

\subsection{Survival through high vibration environments}
The suspension needs to survive the vibratory environments of launch and landing. Significant portions of these room temperature suspensions, as described in this article, are made of fused silica glass. Care must be taken to package these fragile components such they do not suffer damage. The silica fibers are particularly delicate since they will be under stress. Relieving the tension for launch and delivery may be one mitigation, but any damage suffered may still manifest once they are loaded again. Manufacturing them on the moon rather than Earth could be another mitigation, though that requires methods to manufacture them on-cite at the moon, and then weld them to the test mass, assuming similar methods to Advanced LIGO. This article describes using relatively low stress fibers, which will at least help robustness. Vibration tests would be relatively straight forward to investigate mitigation techniques for these high vibration environments.

\subsection{Dust}
Lunar dust contamination of the suspension presents a couple of concerns. Dust on the surface of the optic clearly interferes with the interferometer. It is also unknown how dust may impact the silica fibers. It is possible the breaking stress could be lowered or that they could impact the thermal noise properties. Adequate shielding and processes are required to keep dust away from the suspension. Testing on Earth could be conducted to investigate any degradation in performance.

\subsection{Charging}
Silica is a dielectric and thus subject to charging. Charging can impact the suspension in a number of ways. It could attract dust. Electrostatic forces to nearby hardware may create alignment offsets or create alternate paths for seismic transmission. Charging can impact the performance of electrostatic actuators. Mitigations for charging need to be investigated and could include solutions such as conductive coatings (provided they do not seriously impact thermal noise) or mechanisms such as UV light to discharge the test mass \cite{TMdischarging}.

\subsection{Silica flexure technology development}
To the author's knowledge, no silica anti-springs or silica IP flexures exist. Significant research and development may be needed to realize such technology for two reasons. First, silica is very fragile when under high stress if even microscopic surface defects are present. Second, the required mechanical losses are quite small. While the required losses don't appear from this work to be as strict as the test mass silica fibers, they may still be strict enough to require careful engineering of the flexures.

\subsection{Faults}
It must be assumed that failures of sensors, actuators, and electronics will occur during the operational lifetime of the suspension. These components should be constructed so they are resistant to the moon environment. Redundancy is another common technique for mitigating failures of components. For example, the suspension could be outfitted with extra sensors, actuators, and electronics such that if one or more fail, the suspension is still functional.

\subsection{Thermal fluctuations}
While the temperature on the surface of the moon is not expected to change on very short time scales, it will change significantly from day to night and even during sunny times as the sun moves across the sky over the period of a couple weeks. As these temperature changes occur, the alignment of the suspension will drift as flexure stiffness and geometry evolves. To mitigate these effects and prolong the operational duty cycle of the suspension, measures such as thermal shielding and active temperature control may be desirable. Shielding will also be needed for dust, and it can be placed so the suspension has no view factor of the sun or even the lunar ground. Non-contact heaters can be placed near each spring to actively control suspension low frequency alignment. 

\subsection{Size and delivery}
The suspension must to be delivered to the moon and needs to fit within the volume and mass limitations of a delivery vehicle. Delivery vehicles have not been explored in this work, thus there is work to go in this area. It is however believed that a 1 m, 100 kg suspension could fit within a reasonable vehicle. Multi-stage suspensions that are too long to fit could leave some assembly for the surface of the moon, though this would increase the length and complexity of tasks required in a difficult environment. LILA articles \cite{Cozzumbo, moonseis} assume some astronaut participation, though minimizing complexity at the surface is likely to promote success. 

Lowering suspension mass could be done at the compromise of performance if needed. For multi-stage suspensions, it should be considered whether higher stages can reasonably use lunar regolith as added mass, to reduce the mass required at launch. This in-situ mass utilization would need to trade against concerns such as dust, suspension alignment, thermal noise degradation, and difficulty of tasks required on the moon.

%% file: conclusion.tex
\section{Conclusion}
This paper presents and analyzes three different conceptual room temperature suspension architectures. The different architectures are driven by different seismic noise assumptions. Higher seismic noise motivates more complicated designs. Missions earlier than LILA Horizon, such as LILA Pioneer or the Farside Seismic Suite should provide better limits on the moon's seismic environment to help down-select which architecture is the most appropriate. If the optimistic assumption is justified, a single stage 1 m silica suspension with no vertical isolation may be sufficient. If seismic noise is higher, a longer multi-stage suspension with vertical anti-springs and inverted pendulums may be required. The flexures in the anti-springs and inverted pendulums will likely need to be constructed of silica to reduce thermal noise. Such silica flexures will require technology development. In all cases, utilizing test masses as heavy as can be permitted will reduce the influence of suspension thermal noise. In the case of multiple stages, more mass placed on higher stages will improve performance by lowering the horizontal suspension mode frequencies.

Future work will need to consider the risks associated with realizing such suspensions on the moon: dust, charging, thermal drifts, launch and landing, delivery and installation, faults, and silica flexure technology development. The sensing, actuation, and control system also needs to be developed. All these considerations may further drive suspension architecture choices beyond what has been considered here. 

%% file: appendix.tex
\appendix
\section{Thermal Noise Calculations}
\label{appendix}

\subsection{Material Properties}
\label{subsec:matprop}

Table \ref{tab:materialproperties} lists the material properties required for calculating thermal noise.

%\newpage
\begin{table}[h]
    \caption{Suspension material properties at room temperature \cite{AplusNoise}}
    \label{tab:materialproperties}
        \centering
\begin{tabular}{| l | c | p{0.4\linewidth} |} % Column formatting, 
 \hline
Parameter & Value &  \quad\quad\quad\quad\quad\quad\quad\quad Description \\
 \hline
 
 % Silica
Silica Young's Modulus  $E$ & 72 GPa & Modulus of elasticity of the test mass and the fibers supporting it  \\ \hline
Silica bulk mechanical loss $\phi$ & 4.1e-10 & Internal damping parameter for bulk silica \\ \hline
Silica Dissipation depth &  0.015 m &  Determines the mechanical loss of the surface, important for large surface to volume ratios, as in the fibers \\ \hline
Silica $\beta$ & 1.52e-4 1/K &  Normalized Young's modulus variation with temperature $T$. $\beta=\frac{d}{dT}ln(E)=\frac{1}{E}\frac{dE}{dT}$ \\ \hline
Silica $\alpha$ & 3.9e-7 1/K & Thermal expansion coefficient \\ \hline
Silica density & 2202 kg/m$^3$ & Test mass and fiber density \\ \hline
Silica specific heat & 772 J/Kg/K & Test mass and fiber specific heat \\ \hline
Silica thermal conductivity $K$ & 1.38 W/m/kg & Test mass and fiber thermal conductivity \\ \hline

% C70 Steel for metal wires
C70 Steel Young's Modulus  $E$ & 212 GPa & Metal suspension wire modulus of elasticity of the \\ \hline
C70 Steel bulk mechanical loss $\phi$ & 2e-4 & Internal damping parameter for bulk C70 Steel \\ \hline
C70 Steel $\beta$ & -2.5e-4 1/K &  Normalized Young's modulus variation with temperature $T$. $\beta=\frac{d}{dT}ln(E)=\frac{1}{E}\frac{dE}{dT}$ \\ \hline
C70 Steel $\alpha$ & 12e-6 1/K & Thermal expansion coefficient \\ \hline
C70 Steel density & 7800 kg/m$^3$ & Metal suspension wire density \\ \hline
C70 Steel specific heat & 486 J/Kg/K & Metal suspension wire specific heat \\ \hline
C70 Steel thermal conductivity $K$ & 49 W/m/kg & Metal suspension wire thermal conductivity \\ \hline

% Maraging steel for blade springs
Maraging Steel Young's Modulus  $E$ & 187 GPa & Cantilever spring modulus of elastcity  \\ \hline
Maraging Steel bulk mechanical loss $\phi$ & 1e-4 & Internal damping parameter for bulk Maraging Steel \\ \hline
Maraging Steel $\beta$ & 0 1/K &  Normalized Young's modulus variation with temperature $T$. $\beta=\frac{d}{dT}ln(E)=\frac{1}{E}\frac{dE}{dT}$ \\ \hline
Maraging Steel $\alpha$ & 11e-6 1/K & Cantilever spring thermal expansion coefficient \\ \hline
Maraging Steel density & 7800 kg/m$^3$ & Cantilever spring density \\ \hline
Maraging Steel specific heat & 460 J/Kg/K & Cantilever spring specific heat \\ \hline
Maraging Steel thermal conductivity $K$ & 20 W/m/kg & Cantilever spring thermal conductivity \\ \hline

\end{tabular}
\end{table}

\subsection{The Fluctuation Dissipation Theorem}
\label{subsec:FDT}
The Fluctuation Dissipation Theorem is used in the derivation of thermal noise motion of the test mass resulting from the fact that the temperature is greater than absolute 0, and assuming the suspension is at thermal equilibrium. This theorem in its general form is copied below from Equation 7.12 in \cite{SaulsonBook2ndEd}.

\begin{equation}
x_{therm}^2(f) = \frac{k_BT}{\pi^2f^2}\mathfrak{R}\left(Y(f)\right)
\end{equation}

$x_{therm}$ is the amplitude spectral density of the thermally driven motion of the test mass, $f$ is the frequency in Hz, $T$ is the temperature in Kelvin, and $k_B$ is the Boltzmann constant. $Y$ is the complex valued admittance transfer function from a force on the test mass to the velocity of the test mass. $\mathfrak{R}$ is the real number operator.

Applying this equation to a simple harmonic oscillator, of which a single stage pendulum is an example, results in the equation below, copied (with some minor notation changes) from  Equation 7.39 in \cite{SaulsonBook2ndEd}.

\begin{equation}
x_{therm}^2(f) = \frac{4k_BTk_0\phi}{2\pi f\left[\left(k_0 - m(2\pi f)^2\right)^2 + k_0^2\phi^2 \right]}
\label{eq:single_sus_therm}
\end{equation}

where $k = k_0\left(1 + i\phi\right)$

$k_0$ is the oscillator stiffness, $\phi$ the internal mechanical loss of the material of which the stiffness is constructed, and $m$ the mass. $\phi$ in general could be frequency dependent.

%% file: main.bbl
\providecommand{\newblock}{}
\begin{thebibliography}{10}
\expandafter\ifx\csname url\endcsname\relax
  \def\url#1{{\tt #1}}\fi
\expandafter\ifx\csname urlprefix\endcsname\relax\def\urlprefix{URL }\fi
\providecommand{\eprint}[2][]{\url{#2}}

\bibitem{LILApaper}
Creighton T, Lognonne P, Panning M~P, Trippe J, Quetschke V and Jani K 2025
  {\em arXiv:2508.18437\/} \urlprefix\url{https://arxiv.org/pdf/2508.18437}

\bibitem{LILA_WP}
Jani K, Abernathy M, Berti E, Boschi V, Chakrabarti S, Cocoros A, Conklin J~W,
  Creighton T, Dell'Agnello S, Diels J~C, Eikenberry S, Eubanks T~M, Gill K,
  Grindlay J~E, Izquierdo K, Lee J, Loeb A, Lognonné P, Longo F, Marcano M~P,
  Panning M, do~Vale~Pereira P, Quetschke V, Rahman A, Razzano M, Reed R,
  Shapiro B, Shoemaker D, Smith W, Trippe J, Stryland E~V, Wu W and Yelikar A~B
  2025 {\em arXiv:2508.11631\/}
  \urlprefix\url{https://arxiv.org/pdf/2508.11631}

\bibitem{Cozzumbo}
Cozzumbo A, Mestichelli B, Mirabile M, Paiella L, Tissino J and Harms J 2024
  {\em Philosophical Transactions of the Royal Society A\/} {\bf 382}

\bibitem{moonseis}
Kawamura T and Lognonné P 2010 {\em European Planetary Science Congress\/}
  {\bf 5}

\bibitem{LHOseis}
Harms J 2015 {\em https://dcc.ligo.org/LIGO-T1500224/public\/}

\bibitem{pygwinc}
 2024 pygwinc \urlprefix\url{https://git.ligo.org/gwinc/pygwinc}

\bibitem{Cumming_2012}
Cumming A~V, Bell A~S, Barsotti L, Barton M~A, Cagnoli G, Cook D, Cunningham L,
  Evans M, Hammond G~D, Harry G~M, Heptonstall A, Hough J, Jones R, Kumar R,
  Mittleman R, Robertson N~A, Rowan S, Shapiro B, Strain K~A, Tokmakov K,
  Torrie C and van Veggel A~A 2012 {\em Classical and Quantum Gravity\/} {\bf
  29} 035003 \urlprefix\url{https://doi.org/10.1088/0264-9381/29/3/035003}

\bibitem{AplusFibers}
Lee K~H, Hammond G, Hough J, Jones R, Rowan S and Cumming A 2019 {\em Classical
  and Quantum Gravity\/} {\bf 36} 185018
  \urlprefix\url{https://doi.org/10.1088/1361-6382/ab28bd}

\bibitem{CEnoise}
Consortium C~E 2024 Astrophysical sensitivity
  \urlprefix\url{https://cosmicexplorer.org/sensitivity.html}

\bibitem{ETnoise}
Hild S, Abernathy M, Acernese F, Amaro-Seoane P, Andersson N, Arun K, Barone F,
  Barr B, Barsuglia M, Beker M, Beveridge N, Birindelli S, Bose S, Bosi L,
  Braccini S, Bradaschia C, Bulik T, Calloni E, Cella G, Mottin E~C, Chelkowski
  S, Chincarini A, Clark J, Coccia E, Colacino C, Colas J, Cumming A,
  Cunningham L, Cuoco E, Danilishin S, Danzmann K, De~Salvo R, Dent T, De~Rosa
  R, Di~Fiore L, Di~Virgilio A, Doets M, Fafone V, Falferi P, Flaminio R, Franc
  J, Frasconi F, Freise A, Friedrich D, Fulda P, Gair J, Gemme G, Genin E,
  Gennai A, Giazotto A, Glampedakis K, Gräf C, Granata M, Grote H, Guidi G,
  Gurkovsky A, Hammond G, Hannam M, Harms J, Heinert D, Hendry M, Heng I,
  Hennes E, Hough J, Husa S, Huttner S, Jones G, Khalili F, Kokeyama K,
  Kokkotas K, Krishnan B, Li T~G~F, Lorenzini M, Lück H, Majorana E, Mandel I,
  Mandic V, Mantovani M, Martin I, Michel C, Minenkov Y, Morgado N, Mosca S,
  Mours B, Müller–Ebhardt H, Murray P, Nawrodt R, Nelson J, Oshaughnessy R,
  Ott C~D, Palomba C, Paoli A, Parguez G, Pasqualetti A, Passaquieti R,
  Passuello D, Pinard L, Plastino W, Poggiani R, Popolizio P, Prato M, Punturo
  M, Puppo P, Rabeling D, Rapagnani P, Read J, Regimbau T, Rehbein H, Reid S,
  Ricci F, Richard F, Rocchi A, Rowan S, Rüdiger A, Santamaría L, Sassolas B,
  Sathyaprakash B, Schnabel R, Schwarz C, Seidel P, Sintes A, Somiya K,
  Speirits F, Strain K, Strigin S, Sutton P, Tarabrin S, Thüring A, van~den
  Brand J, van Veggel M, van~den Broeck C, Vecchio A, Veitch J, Vetrano F,
  Vicere A, Vyatchanin S, Willke B, Woan G and Yamamoto K 2011 {\em Classical
  and Quantum Gravity\/} {\bf 28} 094013
  \urlprefix\url{https://doi.org/10.1088/0264-9381/28/9/094013}

\bibitem{AplusNoise}
Barsotti L, McCuller L, Evans M and Fritschel P 2018 {The A+ design curve}
  Internal Technical Document T1800042-v5 LIGO
  \urlprefix\url{https://dcc.ligo.org/LIGO-T1800042/public}

\bibitem{GASprings}
Bertolini A, Cella G, DeSalvo R and Sannibale V 1999 {\em Nuclear Instruments
  and Methods in Physics Research Section A: Accelerators, Spectrometers,
  Detectors and Associated Equipment\/} {\bf 435} 475--483 ISSN 0168-9002
  \urlprefix\url{https://www.sciencedirect.com/science/article/pii/S0168900299005549}

\bibitem{VirgoSus}
Basti A, Boschi V, Chessa P, Dattilo V, Passaquieti R and Ruggi P 2023 {\em
  Nuclear Instruments and Methods in Physics Research Section A: Accelerators,
  Spectrometers, Detectors and Associated Equipment\/} {\bf 1048} 168021 ISSN
  0168-9002
  \urlprefix\url{https://www.sciencedirect.com/science/article/pii/S0168900223000116}

\bibitem{VirgoMAS}
Beccaria M, Bernardini M, Bougleux E, Braccini S, Bradaschia C, Casciano C,
  Cella G, Cuoco E, D'Ambrosio E, {De Carolis} G, {Del Fabbro} R, {De Salvo} R,
  {Di Virgilio} A, Ferrante I, Fidecaro F, Flaminio R, Gaddi A, Gennai A,
  Gennaro G, Giazotto A, Holloway L, {La Penna} P, Losurdo G, Malik S, Mancini
  S, Nicolas J, Palla F, Pan H, Paoletti F, Pasqualetti A, Passuello D,
  Poggiani R, Popolizio P, Raffaelli F, Vicere A, Waharte F and Zhang Z 1997
  {\em Nuclear Instruments and Methods in Physics Research Section A:
  Accelerators, Spectrometers, Detectors and Associated Equipment\/} {\bf 394}
  397--408 ISSN 0168-9002
  \urlprefix\url{https://www.sciencedirect.com/science/article/pii/S016890029700661X}

\bibitem{Edgard}
Bonilla E, Shapiro B and Lantz B 2025 {\em Classical and Quantum Gravity\/}
  {\bf 42} 165007 \urlprefix\url{https://doi.org/10.1088/1361-6382/adf608}

\bibitem{TMdischarging}
Buchman S, Wang S and Saraf S 2025 {\em arXiv:2509.20582\/} (\textit{Preprint}
  \eprint{2509.20582}) \urlprefix\url{https://arxiv.org/abs/2509.20582}

\bibitem{SaulsonBook2ndEd}
Saulson P~R 2017 {\em Fundamentals of Interferometric Gravitational Wave
  Detectors\/} 2nd ed (WORLD SCIENTIFIC) (\textit{Preprint}
  \eprint{https://www.worldscientific.com/doi/pdf/10.1142/10116})
  \urlprefix\url{https://www.worldscientific.com/doi/abs/10.1142/10116}

\end{thebibliography}
